%% file: disaster-tips.tex
\documentclass[10pt,letterpaper]{article}
\input{settings}

\begin{document}
\vspace*{0.2in}

\begin{flushleft}
  {\Large
    \textbf\newline{Twelve Quick Tips for Managing IT Disasters in Small Research Software Teams}
  }
  \newline
  \\
  {Greg Wilson}\textsuperscript{1*}
  \\
  \bigskip
  \textbf{1} Third Bit, Toronto, Ontario, Canada\\
  * Corresponding author, gvwilson@third-bit.com.
\end{flushleft}

In 2025, the US government launched an unprecedented series of attacks on its
own scientific research groups.  A year later \withurl{GitHub}{https://github.com/}
dropped below 90\% availability for the first time, while wildfires in Canada,
France, Spain, and elsewhere forced researchers from the homes and labs.  These
events and others have reminded us just how fragile research computing systems
can be, and that planning for disasters is one of the most effective ways to
prevent them.

This paper is a short guide to disaster planning and recovery for a small
research software team.  The tips assume you are doing everything yourself on
top of your regular job, and that you \emph{aren't} an experienced system
administrator. Some of the tips do require that kind of expertise, but most
research institutions have research computing groups, data librarians, and
environmental health-and-safety offices whose entire job is to help with exactly
these problems. This paper tells you what ``done'' looks like; they can often
provide it.

\section*{Who You Are}

We assume that your team fits one or more of these profiles:

\begin{description}
  
\item[Providing a simple online service]
  Your research group runs a \withurl{Shiny}{https://shiny.posit.co/} app
  or a \withurl{Streamlit}{https://streamlit.io/} dashboard so that collaborators can query data.
  Disasters scenarios include:
  \begin{itemize}
  \item Your cloud provider suspends your account.
  \item The graduate student who deployed the service leaves and no one else has credentials.
  \item Your service depends on a database that another group maintains,
    and that group shuts down on short notice.
  \end{itemize}
  
\item[Publishing a software package]
  Your group releases software to public registries like \withurl{PyPI}{https://pypi.org/} or
  \withurl{CRAN}{https://cran.r-project.org/}. Scenarios include:
  \begin{itemize}
  \item Your PyPI token expires and the person who created it has left.
  \item GitHub suspends your account.
  \item Your institutional web hosting is turned off.
  \end{itemize}
  
\item[Managing data]
  Your group produces datasets in the lab or in the field that are shared with
  collaborators and deposited in repositories.  Scenarios include:
  \begin{itemize}
  \item A lab server's hard drive fails.
  \item A field laptop is stolen and the data on it has not been backed up.
  \item Your data repository announces it is shutting down.
  \end{itemize}
  
\end{description}

These categories overlap: for example, an ecology lab may publish an R package,
run a Shiny dashboard for collaborators, and curate a ten-year field dataset.

\section*{Tip 1: Know your risks.}

Start by making a point-form list of every service and physical asset your team
depends on.  For a small team, this will take about an hour the first time
through, and 15-30 minutes per quarter for review.  A Markdown file in a Git
repository or a Google Doc is the right tool for this.

For our ecology lab, this includes where the R package is published, where the
Shiny dashboard is hosted, where the ten-year field dataset lives, and the
laptops the team works on. For each item, answer two questions:

\begin{enumerate}
  
\item
  How long can you be without it before work stops dead? This is your
  \emph{Recovery Time Objective} (RTO).
  
\item
  How much data can you afford to lose: the last hour's work, the last
  day's, a week's? This is your \emph{Recovery Point Objective} (RPO).
  
\end{enumerate}

Next, identify the single points of failure within your control, such as the
person who knows the deployment process or the credit card that pays for cloud
services. For each, ask what happens if it disappears.

\begin{notebox}[Questions for Support]
  If you have access to research support teams of the kind described in the introduction,
  you may want to ask them these questions.
  \begin{itemize}
  \item Can you help us estimate realistic recovery time and recovery point targets?
  \item Who else shares the systems we depend on, and how would their failure cascade to us?
  \end{itemize}
\end{notebox}

\section*{Tip 2: Make a plan.}

Once you understand your risks, write a plan in a single shared document that
everyone on the team can find in thirty seconds or less. Store the plan in at
least two places that cannot fail simultaneously, such as the shared team drive
\emph{plus} a printed copy in a desk drawer or a PDF saved on every team
member's phone. Pin the link in your team chat, and if you have a physical
office, tape a printed copy to the fridge.

State the disaster declaration criteria in plain language, such as ``our
dashboard has been offline for more than a day'' or ``we have found a virus on a
server''.  Every team member is part of the recovery team, so make sure everyone
has read the plan, understands the criteria, and knows their first action if a
disaster is declared.

Your plan should include a short checklist for each scenario, written as a
numbered series of specific actions. For example, for an online service outage:

\begin{enumerate}
\item Send a message to the \withurl{Signal}{https://signal.org/} group.
\item Log into the database management console and click ``Restore from snapshot''.
\item Check that the dashboard loads and returns data correctly.
\end{enumerate}

\begin{notebox}
  If a step requires knowledge or access that only one person has, you have a
  \emph{lottery-factor} problem (also called a ``bus factor'' problem). Ask
  yourself, ``If Alice wins the lottery and moves to Bali tomorrow, can the rest
  of the team keep things running?''  This is the hardest part of planning: Alice
  probably doesn't recognize all the things only she knows how to do, and no one
  else does either because she has always taken care of it.
\end{notebox}

\begin{notebox}[Questions for Support]
  \begin{itemize}
  \item Will you review our plan and checklists and flag any steps that are wrong or missing?
  \item Where should we keep a copy so your staff can also find it during an incident?
  \item Who is our first point of contact on your team when we declare a disaster?
  \end{itemize}
\end{notebox}

\section*{Tip 3: Back up everything.}

Follow the \emph{3-2-1 rule} \cite{krogh2009}: three copies of every critical
asset, on two different types of media, with one copy off-site. This applies to
data, software releases, electronic lab notebooks, and configuration information.
For our ecology lab, the three copies might be:

\begin{enumerate}
\item
  The working copy on a lab server or shared drive.
\item
  An automated backup. If your institution offers a research computing backup
  service, use it. If not, use a managed database (e.g., \withurl{Amazon RDS}{https://aws.amazon.com/rds/}
  or \withurl{Google Cloud SQL}{https://cloud.google.com/sql}) with automatic
  daily snapshots (no scripting required). For files, GUI tools like
  \withurl{Duplicati}{https://duplicati.com/} or your cloud provider's desktop
  sync app are sufficient.
\item
  An off-site deposit. For datasets, deposit snapshots in an institutional or
  domain repository (\withurl{Dataverse}{https://dataverse.org/},
  \withurl{Dryad}{https://datadryad.org/},
  \withurl{Zenodo}{https://zenodo.org/}, or \withurl{OSF}{https://osf.io/}) that
  provides a DOI. For software, connect Zenodo or
  \withurl{Figshare}{https://figshare.com/} to your repository so every tagged
  release is automatically archived. For lab notebooks, scan or photograph key
  pages quarterly and save them with your other backups.
\end{enumerate}

Back up your source code to at least two different software forges (e.g.,
\withurl{GitLab}{https://gitlab.com/} as well as GitHub) and turn on automatic
mirroring. A \texttt{git push} to a secondary remote is the most cost-effective
one-liner you will ever write, but automatic mirroring is even better.

Back up your configuration as well, such as DNS records, environment variables,
and pipeline definitions. These are small in size but catastrophic to
reconstruct from memory. A single export once a week, stored alongside your
other backups, is probably sufficient.

Finally, \emph{test a full restore from backup} at least once a year: according
to one recent report, almost 20\% of backups fail to restore completely
\cite{backblaze2024}.  And remember that you may need to restore several times
in a real disaster, so check that your restore process is \emph{idempotent},
i.e., that you can run it and then run it again.  A recovery that corrupts data
the second time is as bad as no recovery at all.

\begin{notebox}
  If you can store backups on a separate cloud account with credentials that are
  not stored on any production server, do so. If an attacker compromises your main
  account, they should not get your backups as well.
\end{notebox}

\begin{notebox}[Questions for Support]
  \begin{itemize}
  \item What backup and off-site repository services do you provide?
  \item Can you test our restore process?
  \item Who manages our off-site deposits, and how do we reach them when we need a restore?
  \end{itemize}
\end{notebox}

\section*{Tip 4: Communicate clearly.}

Choose a single communication channel that does not depend on your normal
infrastructure. For example, if you use \withurl{Slack}{https://slack.com/},
agree \emph{in advance} on a Signal group or a phone tree for when Slack is
unavailable.  Make sure the fallback is written in your plan.

For lab and field staff, maintain a printed contact tree: if the server fails at
2:00 a.m., who calls whom? Post it on the lab fridge and save a photo on every
team member's phone.

Keep a printed (or phone-screenshot) contact list with every team member's
mobile number and a personal email address. Do not store this only in a shared
drive that may be affected by an outage.

Pre-write messages and store them with the plan and in your password manager's
shared vault:

\begin{enumerate}
\item
  For the team: ``This is a declared incident: meet in the Signal group.''
\item
  For users of your online service: ``We are investigating an outage and
  will update you within 60 minutes.''
\item
  For collaborators waiting on data: ``We have lost connectivity temporarily.
  Data collected so far is safe; expect a delay of [X] days.''
\item
  For afterward: ``Service is restored and the situation is under control.
  Here is what happened and what we are doing to prevent recurrence.''
\end{enumerate}

Designate one person as the communicator so that everyone else can focus on
fixing the problem. The communicator does not need to be technical: they need to
be calm and reliable. Silence erodes trust faster than bad news, so update
stakeholders at regular intervals (for example, every 60 minutes or at 10:00 am
and 3:00 pm) even if the update is ``still working on it''.

\begin{notebox}[Questions for Support]
  \begin{itemize}
  \item How can we reach your team during an outage?
  \item Is there an institutional status page or mailing list our updates should be mirrored to?
  \item Who on your team needs to receive our incident messages?
  \end{itemize}
\end{notebox}

\section*{Tip 5: Test the plan.}

The most revealing test for a small team is to hand the recovery checklist to
the newest team member, give them a test copy that they can break without
consequences, and ask them to execute the plan without help. Every point where
they get stuck is a documentation or automation gap.

If no one has joined your team recently, gather everyone for an hour-long
walkthrough. Pick the thing that scares you most as ``It's Tuesday morning and
all of Monday's data has disappeared from the database.'' Walk through the
checklist step by step and fix anything that is missing, wrong, or stale then
and there.

Keep track of how long these tests take and compare against your RTO. If
restoring the database takes four hours but your RTO is two hours, you must
either fix the former or adjust the latter.

\begin{notebox}[Questions for Support]
  \begin{itemize}
  \item Can your staff join our next drill so we can find gaps together?
  \item Can you provide a sandbox where we can practice for disasters without affecting production?
  \end{itemize}
\end{notebox}

\section*{Tip 6: Watch for trouble.}

Monitor the things that would get you out of bed at 3:00 a.m. For online
services, free-tier monitoring like \withurl{Uptime Robot}{https://uptimerobot.com/}
and \withurl{Healthchecks.io}{https://healthchecks.io/} require no code and are
infinitely better than nothing.

Check that your dataset's DOI still resolves and that your package is still
installable from its registry. If your software has a ``downloads per month''
badge, a sudden drop to zero may be the first sign that your registry listing is
broken.

Check your cloud bill at least once a month: a misconfigured resource can
generate a catastrophic bill even for a small team. Set budget alerts in your
cloud provider's console at 50\% and 90\% of your normal monthly spend, sent to
at least two people. Most cloud providers offer free-tier budget alerting
adequate for a small team.

\begin{notebox}[Questions for Support]
  \begin{itemize}
  \item What monitoring and alerts do you already run, and can we subscribe to them?
  \item Can you help us set up alerts and uptime checks for our accounts?
  \item Which early-warning metrics would tell us something is about to fail?
  \end{itemize}
\end{notebox}

\section*{Tip 7: Lock down your accounts.}

Enable multi-factor authentication (MFA) on every digital account: this is the
highest-impact security measure you can take \cite{smalls2021}. An authenticator
app or passkey is better than SMS-based MFA, though SMS is better than no MFA at
all.

Use a password manager like \withurl{1Password}{https://1password.com/} or
\withurl{Bitwarden}{https://bitwarden.com/} for the whole team.  Bitwarden's
free tier is often enough for a 3-5 person team. Every shared credential goes in
the password manager: never send credentials over chat or email. Print the
password manager's recovery code and store it somewhere physically safe, because
if you lose access to the password manager you lose access to everything.

If you have a custom domain, know who manages it: institutional IT, a commercial
registrar, or someone on the team. Make sure that person's contact information
is in the disaster plan, that auto-renewal is enabled, and that at least two
people can access the account. Set a quarterly calendar reminder to verify that
auto-renewal is still working.

Make sure at least two people have critical administrative credentials, such as
the cloud root account, the domain registrar login, the payment method for
infrastructure, and the password-manager administrator account.  These
credentials should not be tied to a single phone number or email address.  For
physical access, maintain a list of who has keys to the lab, who knows the alarm
code, and who is authorized to move hardware.

Finally, maintain an offboarding checklist. When someone leaves, revoke their
digital access and collect their physical keys as soon as you can. A disgruntled
former team member with lingering access is trouble you don't need.

\begin{notebox}[Questions for Support]
  \begin{itemize}
  \item Can you help us set up multi-factor authentication?
  \item Which of your team should have access to administrative credentials?
  \item How quickly does your offboarding process revoke access when someone leaves?
  \end{itemize}
\end{notebox}

\section*{Tip 8: Protect your assets.}

Make a simple, explicit policy about where things live, such as, ``patient data
is never copied to a personal laptop'' or ``database dumps are encrypted at
rest''.  If your policy will not fit comfortably on a single page in an 11-point
font, rewrite it until it does.

\begin{notebox}
  You may have obligations you are not aware of under HIPAA, GDPR, or your
  funder's data-management plan.  If you are handling sensitive data, you should
  therefore check whether your institution has an ethics board or data-protection
  office that offers free consultation.
\end{notebox}

Keep laptops and servers patched: enable automatic updates for operating
systems, browsers, and any server-side software. The inconvenience of an
unexpected reboot is less than the inconvenience of ransomware.

For on-premises equipment like a lab server, consider buying a cheap
uninterruptible power supply (UPS) and configure it to trigger a clean shutdown
when the battery reaches 20\%.  Replace the UPS battery every three years: they
have a tendency to degrade silently.

Prepare a short, written response for ransomware and similar breaches. If
someone on the team is contacted by villains, their first action should be to
physically disconnect the affected machine from the network and then call the
designated incident lead. Do not try to negotiate or pay without consulting
legal counsel: most jurisdictions have regulations about ransomware payments.

\begin{notebox}[Questions for Support]
  \begin{itemize}
  \item Who should we talk to about data protection?
  \item Can you confirm our laptops and servers have automatic patching enabled?
  \item If we face ransomware, who is our incident lead and which legal counsel do we call?
  \end{itemize}
\end{notebox}

\section*{Tip 9: Stabilize, then investigate.}

The most important rule of incident response is, ``Stabilize first, investigate
later''.  When something \emph{does} happen, assign clear roles for the duration
of the incident: one person leads the technical response, one person handles
communication using the pre-written messages from Tip 4, and everyone else does
what the leads tell them to.

Create an incident log \emph{as you work}: a shared Google Doc or a thread in
your out-of-band Signal group is good enough.  This log keeps the team in sync
and gives you a record for the post-incident review.  Timestamp every
significant action:

\begin{itemize}
\item 14:32: server outage alarm triggered
\item 14:35: verified that all database requests are returning 404
\item 14:40: began transition to backup server
\end{itemize}

Escalate early. If you are on a cloud provider's support plan, open a ticket the
moment you suspect the problem is on their side. You can cancel your request if
you fix the problem yourself, but you cannot get back the two hours you spent
wrestling with something outside your control.

Finally, conduct a short review within 48 hours of any incident. The goal is to
identify what in the system allowed the incident to happen, not who made a
mistake. Write down one concrete action item and assign it to one person with a
deadline. If you cannot find an outside moderator for the review, have the team
member \emph{least} involved in the incident lead it.

\begin{notebox}[Questions for Support]
  \begin{itemize}
  \item How do we escalate a ticket with our cloud provider(s)?
  \item Who on your team should join our incident response?
  \item Can you act as moderator for our post-incident review?
  \end{itemize}
\end{notebox}

\section*{Tip 10: Count the cost.}

You probably don't have a budget line item titled ``disasters'', but you still
have costs.  Digital costs include cloud backup storage, password-manager
subscription, domain and certificate renewals, managed-database fees, and
repository deposit charges.  More important, but harder to track, are the
personnel costs: time spent on quarterly plan reviews, annual backup-restore
tests and disaster drills, cross-training to reduce lottery factors, and the
hours lost to incident response instead of research.

Calculate what a day of downtime costs you, including things like lost
experiment time, missed paper deadlines, or collaborators who stop trusting
you. That number tells you whether it is worth spending \$50/month on a managed
database or \$200/year on a backup service.

\begin{notebox}[Questions for Support]
  \begin{itemize}
  \item Can you tell us what backup storage, a managed database, and monitoring ought to cost?
  \item Which of those can the institution cover, and which do we have to fund ourselves?
  \end{itemize}
\end{notebox}

\section*{Tip 11: Support your team during the crisis.}

The disasters discussed in the opening are not just logistics problems: each is
a shock to your team and to the people who depend on them.  Team members may be
scared, angry, or ashamed, particularly if they believe they caused the problem.
Handling the human damage is therefore just as important as handling the
technical aspects.

During the acute phase of a disaster, apply the same five principles that the
World Health Organization recommends for crisis responders \cite{who2026}:

\begin{description}
  
\item[Promote safety.]
  Before asking for status updates, confirm that everyone is not in immediate
  personal danger.  Someone whose laptop was stolen may not be thinking clearly
  about data loss because they are still shaken from the theft itself.
  
\item[Calm the environment.]
  Do not add to the panic.  Use a steady voice in your out-of-band chat.  Remind
  the team that the plan exists and that they have rehearsed for this.  Panic is
  contagious, but so is calm.
  
\item[Build self-efficacy.]
  Remind people that they have handled difficult things before.  Give each person
  one concrete task from the checklist so that everyone feels useful rather than
  helpless.
  
\item[Foster connectedness.]
  Use your fallback communication channel to keep the team together.  Isolation
  amplifies distress, so do not let anyone absorb the news alone.
  
\item[Maintain hope.]
  In particular, do not speculate about worst cases before you have confirmed
  them.
  
\end{description}

Protect the person who may have caused the problem.  A team member who
accidentally dropped a table from a database or clicked a phishing link probably
feels worse than anyone else.  Assign them a recovery task, even a small one, so
they can contribute to the fix rather than sitting in the corner imagining their
career is over.  Blame can wait; recovery cannot.

Watch for exhaustion.  Disaster response is an adrenaline sprint followed by a
long slog.  Two people working four hours each will make fewer mistakes than one
person working eight hours straight, so rotate people out of the incident
response when the acute phase is over, even if they insist they are fine.  In
particular, the person who pulled the all-nighter to restore the database should
not also be the person writing the post-mortem the next morning.

\begin{notebox}[Questions for Support]
  \begin{itemize}
  \item What wellbeing or crisis support do you offer, and how do we access it?
  \item Are there any provisions for post-incident leave?
  \item How should crisis preparation and response be factored into annual performance reviews?
  \end{itemize}
\end{notebox}

\section*{Tip 12: Help your team recover after the disaster.}

In Tip 9 said to hold the review within 48 hours. That is for gathering facts
while memory is fresh; a second, more reflective conversation a week or two
later gives people time to process what happened emotionally.  The goal of this
second conversation is not to find root causes but to ask: ``How are we doing?''

If someone on the team caused the disaster, handle it carefully, and make sure
the rest of the team acts that way.  Ask them privately if they are OK.  If you
find yourself wanting to revisit the mistake in every team meeting, you are the
one who hasn't moved on.

Recognize that people recover at different speeds.  One team member may be
joking about it by Tuesday while another is still waking up anxious on Friday.
Both responses are normal.  Check in with each person individually a week after
the incident, and again a month later.  Again, the question is simply: ``How are
you doing with what happened?''

Watch for survivors' guilt.  For example, if your lab's data survived a hacking
intrusion but a neighbouring group lost everything, your team may not want to
celebrate their own recovery when others are still suffering.  Let them talk
about it. If possible, offer practical help to affected colleagues, such as
sharing backup space or simply bringing them coffee.

Finally, celebrate the recovery, not just the disaster. A short thank-you
message that says ``we handled this'' costs nothing and pays dividends the next
time something goes wrong.

\begin{notebox}[Questions for Support]
  \begin{itemize}
  \item What longer-term counselling or support can you offer staff in the weeks after the incident?
  \item Who else in the institution should we inform so they can support our recovery?
  \item Can you help us capture lessons learned so other teams benefit from what happened?
  \end{itemize}
\end{notebox}

\nocite{*}
\bibliography{disaster-tips}

\end{document}

%% file: settings.tex
\usepackage[top=0.85in,left=2.75in,footskip=0.75in]{geometry}

\usepackage{amsmath,amssymb}

\usepackage{changepage}

\usepackage{textcomp,marvosym}

\usepackage{cite}

\usepackage{nameref,hyperref}

\usepackage[right]{lineno}

\usepackage[nopatch=eqnum, expansion=false]{microtype}
\DisableLigatures[f]{encoding = *, family = * }

\usepackage[table]{xcolor}

\usepackage{array}

\newcolumntype{+}{!{\vrule width 2pt}}

\newlength\savedwidth

\raggedright
\usepackage[aboveskip=1pt,labelfont=bf,labelsep=period,justification=raggedright,singlelinecheck=off]{caption}

\makeatletter
\renewcommand{\@biblabel}[1]{\quad#1.}
\makeatother

\usepackage{lastpage,fancyhdr,graphicx}
\usepackage{epstopdf}
\fancyheadoffset[L]{2.25in}
\fancyfootoffset[L]{2.25in}
\usepackage{callouts-box}
\newcommand{\withurl}[2]{{#1}\footnote{{\texttt{#2}}}}